\documentclass[
aps,
pre,
reprint,
twocolumn,
showpacs,
superscriptaddress,
nofootinbib
]{revtex4-2}  

\usepackage{graphicx,color}  
\usepackage{amsmath,amsfonts,amssymb,amsthm}
\usepackage{comment}
\usepackage{soul}
\usepackage{ulem}
\usepackage[colorlinks,citecolor=red,urlcolor=blue,bookmarks=false,hypertexnames=true]{hyperref}

\begin{document}

\normalem

\widetext


\title{
Entropy and canonical  ensemble of hybrid quantum classical systems
}
\date{\today}

\author{J. L. Alonso}
\affiliation{Departamento de F\'{\i}sica Te\'orica, Universidad de Zaragoza,  Campus San Francisco, 50009 Zaragoza (Spain)}
\affiliation{Instituto de Biocomputaci{\'{o}}n y F{\'{\i}}sica de Sistemas Complejos (BIFI), Universidad de Zaragoza,  Edificio I+D, Mariano Esquillor s/n, 50018 Zaragoza (Spain)}
\affiliation{Centro de Astropart{\'{\i}}culas y F{\'{\i}}sica de Altas Energías (CAPA),
Departamento de F\'{\i}sica Te\'orica, Universidad de Zaragoza, Zaragoza 50009, (Spain)}
\author{C. Bouthelier}
\affiliation{Departamento de F\'{\i}sica Te\'orica, Universidad de Zaragoza,  Campus San Francisco, 50009 Zaragoza (Spain)}
\affiliation{Instituto de Biocomputaci{\'{o}}n y F{\'{\i}}sica de Sistemas Complejos (BIFI), Universidad de Zaragoza,  Edificio I+D, Mariano Esquillor s/n, 50018 Zaragoza (Spain)}
\author{A. Castro}
\protect
\affiliation{Departamento de F\'{\i}sica Te\'orica, Universidad de Zaragoza,  Campus San Francisco, 50009 Zaragoza (Spain)}
\affiliation{Instituto de Biocomputaci{\'{o}}n y F{\'{\i}}sica de Sistemas Complejos (BIFI), Universidad de Zaragoza,  Edificio I+D, Mariano Esquillor s/n, 50018 Zaragoza (Spain)}
\affiliation{Fundaci\'on ARAID, Av. de Ranillas 1-D, planta 2ª,
  oficina B, 50018 Zaragoza (Spain)} 
\author{J. Clemente-Gallardo}
\affiliation{Departamento de F\'{\i}sica Te\'orica, Universidad de Zaragoza,  Campus San Francisco, 50009 Zaragoza (Spain)}
\affiliation{Instituto de Biocomputaci{\'{o}}n y F{\'{\i}}sica de Sistemas Complejos (BIFI), Universidad de Zaragoza,  Edificio I+D, Mariano Esquillor s/n, 50018 Zaragoza (Spain)}
\affiliation{Centro de Astropart{\'{\i}}culas y F{\'{\i}}sica de Altas Energías (CAPA),
Departamento de F\'{\i}sica Te\'orica, Universidad de Zaragoza, Zaragoza 50009, (Spain)}
\author{J. A. Jover-Galtier}
\affiliation{Departamento de F\'{\i}sica Te\'orica, Universidad de Zaragoza,  Campus San Francisco, 50009 Zaragoza (Spain)}
\affiliation{Instituto de Biocomputaci{\'{o}}n y F{\'{\i}}sica de Sistemas Complejos (BIFI), Universidad de Zaragoza,  Edificio I+D, Mariano Esquillor s/n, 50018 Zaragoza (Spain)}
\affiliation{Centro Universitario de la Defensa de Zaragoza, Academia General Militar, 50090 Zaragoza (Spain)}

\begin{abstract}

    In this { work} we generalize and combine Gibbs and von Neumann
approaches to build, for the first time,  a rigorous definition of 
entropy for hybrid quantum-classical systems. The resulting
function coincides with the two cases above when the suitable limits
are considered.
Then, we apply  the MaxEnt principle for this hybrid
entropy function and obtain the natural
candidate for the Hybrid  Canonical Ensemble (HCE). We prove
that  the suitable classical  and quantum limits of the
HCE coincide with the usual classical and quantum canonical ensembles
since the whole scheme admits both limits, thus showing that the MaxEnt principle is applicable and consistent for hybrid systems. 

\end{abstract}

\pacs{}
\maketitle

\section{Introduction}
Hybrid quantum-classical (QC) systems are the natural approximation
to those quantum systems containing some degrees of freedom that can
be well approximated as classical variables.  This possibility arises
when there are two different energy or mass scales, as it happens, for
instance, in molecular and condensed matter systems where the nuclei
are heavy and slow, while the electrons are light and
fast. Hybrid
models have also been proposed to explain the measurement process~
\cite{Diosi2014,Buric2013}: the measurement device is modeled
as  a classical system
coupled to the quantum system to be measured.  In field theory, hybrid
quantum-classical systems have also been considered as candidates
to describe quantum matter fields interacting with a
(classical) gravitational field, as a semiclassical approximation or
even a fundamental theory (see \cite{Martin-Dussaud2019,Tilloy2019}). 

The correct mathematical formalism for the dynamics and statistics of
these hybrid models is not obvious.  Two different points of view can
be taken. On the one hand, a practical one: the construction of a
hybrid theory that approximates, as closely as possible, the full
quantum dynamics of the problem. Such methods can be applied to a very
large array of problems in condensed matter and molecular physics and
chemistry, as non-adiabatic processes play a fundamental
role~\cite{Tully1998b, Yonehara2012,Tavernelli2015, CrespoOtero2018, Curchod2018}.  On the other hand, a fundamental,
theoretical point of view: the construction of a mathematically and
physically consistent theory for hybrid systems, according to 
{different demands of consistency}~\cite{Prezhdo1997,Kisil2005,Prezhdo2006,Salcedo2007,Agostini2007,Kisil2010,Agostini2010,Hall2008,Buric2013b,Peres2001,Terno2006,Salcedo1996,Gil2017,Caro1999,Diosi2014,Elze2012,Aleksandrov1981,Kapral1999,Alonso2011}, independently of how well
it may approximate the full quantum dynamics. This
second  approach  is compulsory when the full quantum dynamics is not known, as in the
  case of a system of quantum matter fields interacting with gravity.
In any case, it is not clear what is the best possible dynamics from
any of those two points of view. {  Here, we assume the second one,
  and add to the discussion  on the construction
of a mathematically consistent and physically motivated
hybrid theory.}

The focus of this { work} is on the statistical mechanics of
hybrid systems, \emph{regardless of the dynamics} chosen for
their description.  In particular, we consider  two
   open
   questions:  first, {  what is the correct definition of the entropy of a hybrid system? And then, given this definition, }
 can we use the MaxEnt formalism and obtain the
  canonical ensemble of a hybrid system, as we do for classical or
  quantum ones? Apparently, these are purely fundamental questions, but their
  answers are crucial for
many applications, in particular {  for} the ab initio modeling of molecules and materials
and their numerical simulation methods at finite temperature (for
example, see \cite{Alavi1994,Grumbach1994,Silvestrelli1996, Ji2013, Ruter2014,Karasiev2014}). { We  determine, in a simple
 way, the equilibrium ensemble that  the numerical methods
must reproduce and the entropy function they must consider.}

The structure of the {  paper is as follows.
In Section \ref{Sec:II} we will first discuss the proper definition of
the hybrid entropy function.
  Then, in Section \ref{Sec:III} we will derive the 
hybrid canonical ensemble (HCE) as the one that maximizes this
entropy, subject to the constraint of a given expectation value for
the energy (MaxEnt principle).
 The resulting ensemble had been perhaps implicitly assumed before,
but few times explicitly spelled, and never, to our knowledge, derived
from the general principle of entropy maximization.  {\color{black} We
  will also briefly discuss some relevant properties of the resulting ensemble.}
Finally, in Section \ref{Sec:V} we will summarize our main conclusions.}
\mbox{}

\section{The entropy of a hybrid QC system.}
\label{Sec:II} 
A correct
statistical mechanical definition of any system departs from the
definition of a sample space: a set of statistically
independent states, i.e.  a basis of mutually exclusive events (MEE), which
can be unequivocally characterized by the results of an experiment.
Let us start by recalling the basic definitions in the { purely}-classical
or { purely}-quantum cases.

In classical systems, a basis of MEEs is simply the phase space
$\mathcal{M}_C$, the set of all positions and momenta of the classical
particles: $\mathcal{M}_C=\left\lbrace (Q,P)\mid Q\in\mathbb{R}^n,\;
P\in\mathbb{R}^n\right\rbrace $, where $n$ is the number of classical
degrees of freedom. Any point in this phase space 
defines an exclusive event from any other event. Observables are real
functions on this $\mathcal{M}_C$. Statistical
mechanics for classical systems can then be described by using
\emph{ensembles} on this phase space, i.e. (generalized)\footnote{We introduce the adjective \textit{generalized} to refer to the set of generalized functions (or distributions) and include, for instance, Dirac delta functions.} probability distribution functions (PDFs) $F_{\rm C}$ 

In quantum systems, the states  are rays of a Hilbert space
$\mathcal{H}$, i.e. the analogous to the classical phase space is the
projective space, $\mathcal{M}_Q = \mathcal{PH}$. {  We will represent its points as the projectors on 1-dimensional subspaces of the Hilbert space $\hat \rho_\psi=\frac{|\psi\rangle \langle\psi|}{\langle\psi|\psi\rangle }$, with $|\psi\rangle \in \mathcal{H}\setminus \{\vec 0\}$.} Even though all of
the states in $\mathcal{M}_Q$ are physically legitimate, they are not
mutually exclusive. Indeed,  if the system has been measured to be, with
probability one, in a state { $\hat \rho_{\psi_1}$}, the probability of measuring it
to be in other state { $\hat \rho_{\psi_2}$} is not zero, unless they are orthogonal:
{ $\hat \rho_{\psi_1},\hat \rho_{\psi_2}$} are MEE only if $\langle\psi_1\mid\psi_2\rangle=0$. As
a consequence, considering { generalized} probability density functions
$F_{\rm Q}$ over the Hilbert space (or over
the projective space of rays) to define ensembles, following the
classical analogy, results in over-counting the same outcome for a
hypothetical experiment in a non-trivial way. One way to see this
clearly is that many different $F_{\rm Q}$ can correspond to exactly the
same ensemble (i.e.  they are physically indistinguishable). The
correct way to get a sample space of MEEs is therefore considering a
basis of orthogonal events. From this idea, von Neumann~\cite{vonNeumann1955}
derived the density matrix formalism, which contains all the
physically relevant statistically non-redundant information in a
compact way. A density matrix can be obtained from a PDF $F_{\rm Q}$
in the quantum { state}  space as:
{ \begin{equation}
\hat{\rho}[F_{\rm Q}] =\int{\rm d}\mu_Q(\hat \rho_\psi)F_{\rm Q}(\hat \rho_\psi)\hat \rho_\psi\,,
\end{equation}
where we represent by $d\mu_Q$ the volume element on $\mathcal{M}_Q$. Analogously, in the following, we will represent by $d\mu_C$ the volume element on $\mathcal{M}_C$.
}


{ 
We move on now to QC theories. Despite the various proposals
{\color{black} referenced} above, one can perhaps establish a common denominator.
The classical part is described by a set of position $Q \in \mathbb{R}^n$
and momenta $P \in \mathbb{R}^n$ variables, that we will hereafter
collectively group as $\xi = (Q, P)$. The quantum part is described by a
complex Hilbert space $\mathcal{H}$. Observables are Hermitian operators
on $\mathcal{H}$, and they may depend parametrically on the classical
variables, $\hat{A}(\xi) : \mathcal{H} \to \mathcal{H}$.  Those
observables defined on the classical subsystem  are just
$\xi$-functions times the identity, i.e. $\hat{A}(\xi) =
A(\xi)\hat{I}$; those observables defined on the quantum subsystem
only are operators that lack the $\xi$-dependence.

}

We are going to consider two different approaches to the definition of the entropy, one based on the usual approach to classical systems, and another one inspired by the quantum case.

{ \subsection{A Gibbs-entropy for hybrid systems?}
}

{ The formal similarities of one of the best known hybrid dynamical models, Ehrenfest dynamics, with the classical one (see \cite{Alonso2011,Buric2013b} for details) {  may lead to consider hybrid systems as formally closer to classical than to quantum dynamics}. Indeed, Ehrenfest dynamics can be given a Hamiltonian {  structure} (see \cite{Bornemann1996, Alonso2011}) in terms of 
\begin{itemize}
 \item a Hamiltonian function constructed as
{ \begin{equation}
 \label{eq:hybridHamiltonian}
 f_H(\xi, \hat \rho_\psi)=\mathrm{Tr}(\hat H(\xi) \hat \rho_\psi)=\frac{\langle \psi| \hat H(\xi)\psi\rangle}{\langle\psi|\psi\rangle },
\end{equation} 
}\item and a Poisson bracket obtained as the combination of the Poisson bracket of Classical Mechanics and the canonical Poisson bracket of quantum systems (see \cite{Kibble1979,Heslot1985}).

\end{itemize}

This fact makes {  Ehrenfest dynamical description of hybrid system formally analogous to a classical Hamiltonian dynamical system}. When considering the definition of hybrid statistical systems, we can then consider a  hybrid   { (generalized)} PDF $F_{\rm H}$ defined over the hybrid phase space $\mathcal{M}_H = \mathcal{M}_C \times \mathcal{M}_Q$, in an analogous manner to the definition of classical statistical systems. The Hamiltonian nature of the dynamics allows to  define a Liouville equation for  $F_{\rm H}$ in a straightforward manner (see \cite{Alonso2011,   Buric2013b}). 

Within that framework, it is also tempting to borrow the notion of entropy from Classical Statistical Mechanics and define a Gibbs-like function associated with the density function $F_{\rm H}$ in the form:
{ \begin{equation}
 \label{eq:GibbsEntropy}
 S_{\rm G}[F_{\rm H}]=-k_B \int_{\mathcal{M}_H} d\mu_H(\xi, \rho_\psi) F_{\rm H}(\xi,\rho_\psi) \log \left( F_{\rm H}(\xi,\rho_\psi) \right),
\end{equation} 
where $k_B$ represents the Boltzman costant and $d\mu_H$ represents the volume element on $\mathcal{M}_H$ which can be written in terms of the classical and quantum volume elements as $d\mu_H=d\mu_C\wedge d\mu_Q$.
}

Notice that this entropy function is well defined for classical systems, where the points of phase-space correspond to mutually exclusive events. Therefore, when considering $S_{\rm G}$ we are
adding all points of the phase space $\mathcal{M}_H$ as if they were mutually exclusive. Thus we treat them as  classical statistical systems, where being at a given point in phase space excludes the possibility of being at a different point.  
{Hence, we are not weighting correctly the quantum subsystems from the physical point of view, ruining the function ability to measure physical information for the hybrid system.}

Despite this fact, this entropy function has been implicitly assumed several times when considering hybrid or even purely-quantum statistical systems (see \cite{Brody1998, JonaLasinio2006,Alonso2011, Campisi2013a}), when defining the so called Schr\"odinger-Gibbs (SG) ensemble or the corresponding Schr\"odinger microcanonical ensemble. {  Thus, SG represents a canonical ensemble where the probability density is written by assigning} to each state the Gibbs weight associated with the expectation value of the Hamiltonian, instead of the operator itself. 
But the bad physical properties of $S_{\rm G}$ lead to very strange and un-physical properties for the corresponding thermodynamic functions.
In particular, this was the case when the Schr\"odinger-Gibbs ensemble was analyzed in \cite{Alonso2015}.  Nonetheless, notice that $S_{\rm G}$ {is a mathematically consistent entropy function, despite the unphysical properties  of the Statistical Mechanics it defines.}

\subsection{ Gibbs-von Neumann entropy}
From our analysis above, it is clear that the straightforward extension of Gibbs classical entropy function to hybrid systems leads to inconsistencies because the points of hybrid phase space do not define mutually exclusive events as the classical phase space points do.}
In order to do statistical mechanics in a consistent way with the nature of its quantum subsystem, one must {  reconsider the notion of mutually exclusive events, and combine the classical and the quantum notions of MEE}. The combined hybrid phase space is 
  now $\mathcal{M}_H = \mathcal{M}_C
\times \mathcal{M}_Q$. But, we must consider that two hybrid
states $(\xi_1, \hat \rho_{\psi_1}), (\xi_2, \hat \rho_{\psi_2})\in \mathcal{M}_H$ represent
MEEs if and only if $\xi_1\neq \xi_2$ or $\langle \psi_1|\psi_2 \rangle=0$.

{ The next step is} to define a probability distribution on the set of MEEs of $\mathcal{M}_H$. {  Following von Neumann idea and the mathematical construction of Gleason theorem \cite{Gleason1957}, we can build a hybrid density matrix to represent the hybrid probability in a consistent way.  }
As the physical properties of the hybrid system, in general,
combine the states of $\mathcal{M}_C$ and $\mathcal{M}_Q$ (for
instance, the total energy of the system), we cannot
expect both sets to be independent from the probabilistic point of
view. Nonetheless, we can assume that we can simultaneously measure
any classical observable and any hybrid observable of the form $\hat A(\xi)$.
{ This fact permits to define the conditional probabilities $p(a|\xi)$: the probability of measuring an eigenvalue a of operator $\hat A(\xi)$, given that the classical subsystem is at state $\xi\in \mathcal{M}_C$. The probabilities associated  to the hybrid measurement can then be decomposed into the marginal probability associated to the classical phase space, $F_C(\xi)$, and the conditional probabilities
associated to the measurement of $\hat{A}(\xi)$, given $\xi$:
\begin{equation}
  \label{eq:1}
 p(a, \xi)=F_{\rm C}(\xi)p(a|\xi).
\end{equation}
}

For these quantum conditional probabilities { $p(a|\xi)$}, all the requirements
of Gleason's theorem \cite{Gleason1957} apply, and one may therefore
define, at each $\xi$-point, a density matrix $\hat{\rho}^\xi$. It provides
the probabilities of measuring an eigenvalue $a$ of observable
$\hat{A}(\xi)$, given $\xi$, through  the usual  Born rule:
$p(a\vert \xi) = {\rm Tr} [\hat{\rho}^\xi \hat{\pi}_a(\xi)]$, where $\hat{\pi}_a(\xi)$ is
the projector onto the eigen-subspace associated to $a$. {  From this, we can} define the \emph{hybrid density matrix} {  as the $\xi$--dependent matrix: 

\begin{equation}
  \label{eq:6}
  \hat \rho(\xi)=F_{\rm C}(\xi)\hat \rho^\xi,
\end{equation}
 such that $p(a, \xi)=\mathrm{Tr}(\hat \rho(\xi)\hat \pi(a))$}.
{   Notice that, strictly speaking,  Gleason
    theorem ensures  the existence and uniqueness of the density matrix
    $\hat \rho^\xi$ only for Hilbert spaces of
    dimension at least 3. 
However, the recent developments based on positive-operator-valued measures (POVM)
(see for instance \cite{Busch2003,Caves2004}) allow to prove a
more general formulation of Gleason theorem  for quantum states which
is valid in  dimension 2,  but in that case the  construction is not 
based on orthogonality of the rank-one projectors but on a more
global set of \textit{effects}.
}

 In conclusion, the probability distribution on the set of MEEs of hybrid
states can be written as a family of quantum density operators
parameterized by the classical degrees of freedom, $\hat{\rho}(\xi)$.
For each $\xi$, $\hat{\rho}(\xi)$ is a self-adjoint and
non-negative  operator, which is normalized on the full hybrid sample space:
\begin{equation}
\int_{\mathcal{M}_C}\!\!{\rm d}\mu_C(\xi) \mathrm{Tr}(\hat{\rho}(\xi)) = 1\, .
\end{equation}
This is an immediate consequence of the normalization of $F_{\rm
  C}(\xi)=\mathrm{Tr}\hat \rho(\xi)$ { $\left (\int_{\mathcal{M}_C}d\mu_C(\xi)F_{\rm C}(\xi)=1 \right )$} and  of  $\hat \rho^\xi$ $(\mathrm{Tr}\hat{\rho}^\xi =  1)$.
Given a hybrid state determined by the classical point $\xi$ (which has
probability $\mathrm{Tr}\hat \rho(\xi)$), and a
quantum state represented by the projector $\hat{\pi}$, the
probability of measuring the system to be in that state is given by
$\mathrm{Tr}(\hat{\rho}(\xi)\hat{\pi})$.
These $\xi-$dependent density matrices have already been used
before, for example by Aleksandrov~\cite{Aleksandrov1981}, or obtained
by taking the partial classical limit in the Wigner transformation of
the full quantum density matrix, in the quantum-classical Liouville
equation method~\cite{Kapral1999}.

Let us consider now how to define the entropy of these hybrid states.
For any bivariate distribution $p(x,y)$  of two sets of random
variables ($X$, $Y$), the entropy $S(p)$ decomposes as
\begin{equation}
S(p) = S(p_X) + \sum_x p_X(x)S(p_{Y\vert x})\, ,
\end{equation}
where $p_X(x) = \sum_y p(x,y)$ is the marginal distribution of $X$, and
$p_{Y|x}$ is the conditional probability of $Y$ given $x$. This
general result must be  applicable to the decompositions \eqref{eq:1}
and \eqref{eq:6} .
{  Therefore, the entropy of the hybrid system must be equal to the sum
of the (classical) entropy ($S_C$)  of the marginal classical distribution $F_{\rm C}(\xi)$ and the average, over $F_{\rm C}(\xi)$, of the (von Neumann) entropy associated to the  conditional probability $\rho^\xi$, i.e.:
\begin{multline}
\label{eq:descomp}
S[\hat{\rho}(\xi)] = \overbrace{-k_B \int_{\mathcal{M}_C}\!\! {\rm
    d}\mu_C(\xi) F_{\rm C}(\xi) \log(F_{\rm C}(\xi))}^{S_C(F_{\rm C})} + 
\\
\int_{\mathcal{M}_C}\!\! {\rm d}\mu_C(\xi)  F_{\rm C}(\xi) 
\underbrace{\left[-k_B
\mathrm{Tr} \left ( \hat \rho^\xi \log \hat \rho^\xi\right )
\right]}_{S_{\rm vN}(\hat \rho^\xi)}
\end{multline}
}
 It is immediate then to rewrite this as:
\begin{equation}
\label{eq:hybrid_entropy}
	S[\hat{\rho}(\xi)]=-k_B\int_{\mathcal{M}_C}\!\!{\rm d}\mu_C(\xi)\; \mathrm{Tr}\left ( \hat
          \rho(\xi)\log \hat \rho(\xi)\right  ),
\end{equation}
\emph{which is our proposal for the hybrid QC entropy.} 
  To the best of our knowledge, this is the first rigorous proposal of an entropy 
  function for a hybrid quantum-classical system.
{  If the classical subsystem is pure, 
(i.e. $F_{\rm C}(\xi)=\delta(\xi-\xi_0)$)  the classical entropy vanishes and} the entropy above reduces to von
Neumann entropy. {  Analogously, when the quantum state is pure and
independent of the classical state,  the von Neumann entropy of $\rho^\xi$ vanishes, and the expression} above reduces to
the classical entropy function. Therefore,
the entropy function (\ref{eq:hybrid_entropy}) 
combines the classical and quantum information in a consistent way,
and has the correct classical and quantum limits.

\mbox{}

\section{The MaxEnt principle for hybrid QC systems.}  
\label{Sec:III} 

\subsection{ MaxEnt principle for the hybrid entropy function}
The maximum
entropy principle is one of the standard procedures to derive the
canonical ensemble at both the classical or the quantum
level. Firstly, one must assume that the system is in equilibrium.
Then, one can find the canonical ensemble as the solution of the
MaxEnt problem: given a certain thermodynamic system and an entropy
function $S$, find the equilibrium ensemble which maximizes $S$
 among those with a fixed value of the {\color{black} average  energy
   $E=\langle  \hat H (\xi)\rangle$}.

In the following, we will prove that \emph{the canonical ensemble that
results of this maximization, for the hybrid case, is given by}:
\begin{align}
\label{eq:hce} 
\hat \rho_{\rm HCE}(\xi) &= \dfrac{e^{-\beta \hat 
H(\xi)}}{Z_{\rm HCE}(\beta)}
\\
Z_{\rm HCE}(\beta) &= 
\int_{\mathcal{M}_C}\!\!{\rm d}\mu_C(\xi)\;\mathrm{Tr}(e^{-\beta \hat H(\xi)})
\end{align}
where $\hat{H}(\xi)$ is the Hamiltonian (typically decomposed into a classical and
a quantum part,  as $f_{H}^c(\xi)\hat{I} + \hat{H}_Q(\xi)$), $Z_{\rm HCE}(\beta)$ is the
partition function, and $\beta$ is a constant, determined by the
choice of $E$, that is used to define the (inverse of the)
temperature. Note that this ensemble had been perhaps implicitly assumed before, but
seldom explicitly written~\footnote{
For example, it was given in Ref.~\cite{Mauri1993}, where it was claimed to be
the partial classical limit of the fully quantum canonical ensemble.
It was also presented as the zero-th order term in a classical-limit expansion of the
partial Wigner transformation of the quantum canonical ensemble in Ref.~\cite{Kapral1999}. Finally, 
in footnote 30 of Ref.~\cite{Alonso2012c}, some of the current authors already hinted, without
proof, the result demonstrated here.
} and, to our knowledge, never derived.  Notice that the
  orthogonal projectors of its spectral decomposition coincide with those of the
  adiabatic basis.

The problem can be addressed as a constrained optimization problem:
find the  density matrix that maximizes $S$ in
Eq.~(\ref{eq:hybrid_entropy}), subject to  the constraints:
\begin{align}
 \label{eq:N}
C_N[\hat{\rho}(\xi)] :&=\int_{\mathcal{M}_C}\!\!{\rm d}\mu_C(\xi) \mathrm{Tr}(\hat{\rho}(\xi))-1 = 0\,,
\\
 \label{eq:U1} 
C_E[\hat{\rho}(\xi)] :&=\int_{\mathcal{M}_C}\!\!{\rm d}\mu_C(\xi) \mathrm{Tr}(\hat{\rho}(\xi)\hat{H}(\xi))-E = 0\,.
\end{align}
These
can be incorporated  \textit{via} Lagrange multipliers, defining the full optimization functional to be:
\begin{equation}
\label{eq:J} 
\mathcal{S}:=S-\lambda_N C_N-\lambda_E C_E.
\end{equation}

Without loss of generality, let us work in the ($\xi$-dependent) basis of
eigenstates of the Hamiltonian (the \emph{adiabatic} basis).
First, we will consider the optimization over a reduced set of density matrices:
those which are diagonal in this adiabatic basis.
 The terms in Eq.~(\ref{eq:J}) then read:
\begin{align}
S[\lbrace\rho_{ii}\rbrace] &= -k_B \int_{\mathcal{M}_C}\!\!{\rm d}\mu_C(\xi)\; \sum_i \rho_{ii}(\xi)\log(\rho_{ii}(\xi)\,,
\\
C_N[\lbrace\rho_{ii}\rbrace] &= \int_{\mathcal{M}_C}\!\!{\rm d}\mu_C(\xi)\; \sum_i \rho_{ii}(\xi) - 1
\\
C_E[\lbrace\rho_{ii}\rbrace] &= \int_{\mathcal{M}_C}\!\!{\rm d}\mu_C(\xi)\; \sum_i H_i(\xi) \rho_{ii}(\xi) - E
\end{align}
  Taking derivatives and setting them to zero leads
immediately to
\begin{equation}
\label{eq:hcediagonal}
\rho_{ii}(\xi) = Z_{\rm HCE}(\beta)^{-1}e^{-\beta H_i(\xi)}\, ,
\end{equation}
where $\beta=\frac{\lambda_E}{k_B}$.

We consider now a general density matrix $\hat{\tilde{\rho}}(\xi)$,
whose non-diagonal elements may be non-zero, fulfilling the two
constraints \eqref{eq:N} and \eqref{eq:U1}.  Since it is Hermitian with
non-negative eigenvalues, it satisfies Klein's lemma~\cite{Klein1931}:
\begin{equation}
-\mathrm{Tr}(\hat{\tilde{\rho}}(\xi)\log(\hat{\tilde{\rho}}(\xi))\leq -\sum_i \tilde{\rho}_{ii}(\xi) \log(\tilde{\rho}_{ii}(\xi))
\end{equation}
where $\tilde{\rho}_{ii}(\xi)$ are its diagonal elements (the equality
only holds if it is actually diagonal).  As
  the constraints (\ref{eq:N}) and (\ref{eq:U1})
in the adiabatic basis only depend on the diagonal elements of $\hat
\rho(\xi)$, we may conclude  {\color{black} that for
any non-diagonal density matrix that fulfills the constraints there
exists a diagonal one (defined to be the one whose diagonal entries
are the same) that also fulfills the constraints and has a larger
entropy.} The global maximum, therefore, has to 
be found among the  diagonal ones,
and is the one given  in Eq.~(\ref{eq:hcediagonal}). This concludes the proof.

\subsection{Properties of the HCE}
Let us now check that the ensemble thus defined fulfills
some very natural requirements:
\begin{itemize}

\item \emph{Additivity}.  If two systems are in the canonical ensemble
  equilibrium at the same temperature, they must also be at
  equilibrium when we consider them to form a single systems with two
  (independent) subsystems. Extensive variables as the energy and
  entropy must be additive.

This can be proven for the HCE in the following way. If
$\hat{H}_1(\xi_1)$ and $\hat{H}_2(\xi_2)$ are the Hamiltonians of both
systems, the combined one is:
\begin{equation}
\label{eq:total_hamiltonian} 
\hat{H}(\xi)=\hat{H}_1(\xi_1)\otimes \hat{\mathbb{I}}_2+\hat{\mathbb{I}}_1\otimes \hat{H}_2(\xi_2),
\end{equation}
{ where $\xi=(\xi_1, \xi_2)$}.

As the two terms of \eqref{eq:total_hamiltonian} trivially commute,
\begin{equation}
e^{-\beta \hat H(\xi)} = e^{-\beta \hat{H}_1(\xi_1)}\otimes e^{-\beta \hat{H}_1(\xi_2)}\,,
\end{equation}
and because of this, 
\begin{multline}
\int_{\mathcal{M}_{C_1}\times \mathcal{M}_{C_2}}\, d\mu_C(\xi_1,\xi_2)  \mathrm{Tr}\,e^{-\beta \hat H(\xi)} =
\\  \int_{\mathcal{M}_{C_1}}d\mu_C(\xi_1)  \mathrm{Tr} \, e^{-\beta \hat H(\xi_1)}
\int_{\mathcal{M}_{C_2}}d\mu_C(\xi_2)  \mathrm{Tr}\, e^{-\beta \hat H(\xi_2)}
\end{multline}

Thus we can just write
\begin{equation}
\hat\rho(\xi)=\hat \rho_1(\xi_1)\otimes \hat \rho_2(\xi_2)
\end{equation}

This factorization of $\hat \rho(\xi)$ immediately implies the additivity
of the internal energy \eqref{eq:U1} and of the entropy
\eqref{eq:hybrid_entropy}.

\item The classical canonical ensemble, which maximizes Gibbs entropy,
  is recovered when only one quantum energy state exists.

\item The quantum canonical ensemble, which maximizes von Neumann
  entropy, is recovered when only one classical point is allowed.

\item If the QC coupling is turned off (the quantum
  Hamiltonian $\hat{H}_Q$ is independent of the classical
  variables and \textit{vice versa}), the HCE becomes the product of the
  classical and quantum canonical ensembles, which maximize the sum of
  their respective entropies independently.
\end{itemize}

\mbox{}

\subsection{Dynamics.}
\label{Sec:IV} 
Another extra condition that an equilibrium ensemble
must obviously verify is missing in the previous list: stationarity
under the dynamics  of the micro-states. However, up to
now we have disregarded  the
dynamics, and derived the  canonical
ensemble from very broad assumptions, freed from dynamical
arguments. { The dynamics is  neither relevant for the
  definition of the entropy function nor  affects directly  the solution
  of the MaxEnt condition.  For instance, ${\rm MaxS_G}$ defines the MaxEnt solution for the entropy function $S_{\rm G}$ (Eq. \eqref{eq:GibbsEntropy}), independently of the dynamics of the microstates we consider. The existence of dynamics having it as an equilibrium point would be an extra requirement for the definition of a thermodynamical ensemble. 
  
  
On the other hand, we also proved above that the MaxEnt solution of the true hybrid entropy function \eqref{eq:hybrid_entropy} is the HCE. This implies that the only possible ensemble which can be considered to represent the canonical ensemble of a hybrid system is the HCE. Is there  a dynamics that makes it also stationary? 
Trivially, the commutator with $\hat H(\xi)$ (i.e. a
generalized von Neumann equation) does, but
many others may also be possible. We will analyze this issue in a
forthcoming publication. 

}

\section{Conclusions}
\label{Sec:V} 

It has been the purpose of this paper to shed some light into the issue
of the entropy and the canonical equilibrium expression for hybrid
systems.  We have first discussed the definition for the entropy of an
ensemble of hybrid systems. We have done it by making very general
assumptions on the hybrid theory, but without any consideration for
the particular dynamics. { We have considered two different alternatives, one based on probability densities on the hybrid phase space and another based on projectors and the notion of hybrid mutually exclusive events. The first case leads to a Gibbs-like function which treats the hybrid system as a direct analogue of a classical system. We have { shown how} that entropy function assigns the wrong weight to hybrid events and because of this fails to produce a physically meaningful Thermodynamics. The second proposal  departs from the
information-theory definition of entropy, and carefully considers the
principle of mutually exclusive events. The resulting hybrid entropy function weights correctly the hybrid exclusive events and defines a physically consistent thermodynamical entropy.  

Then, we have derived the HCE
as the one that fulfills the MaxEnt principle with respect to the hybrid entropy function},  using it for the 
first time for hybrid quantum-classical
systems. Furthermore, we verified that {the  HCE} reproduces the
classical and quantum cases when the suitable limits are considered. {
  Hence, we can claim that the MaxEnt principle is applicable and
  consistent for hybrid {\color{black} quantum-classical} systems}.


\acknowledgments{
The authors would like to thank Profs. 
Floria and Zueco for their very useful suggestions.  Partial
financial support by MINECO Grant FIS2017-82426-P  is
acknowledged. C. B.  acknowledges financial support by Gobierno de
Aragón through the grant defined in ORDEN IIU/1408/2018. 
}



\begin{thebibliography}{53}%
\makeatletter
\providecommand \@ifxundefined [1]{%
 \@ifx{#1\undefined}
}%
\providecommand \@ifnum [1]{%
 \ifnum #1\expandafter \@firstoftwo
 \else \expandafter \@secondoftwo
 \fi
}%
\providecommand \@ifx [1]{%
 \ifx #1\expandafter \@firstoftwo
 \else \expandafter \@secondoftwo
 \fi
}%
\providecommand \natexlab [1]{#1}%
\providecommand \enquote  [1]{``#1''}%
\providecommand \bibnamefont  [1]{#1}%
\providecommand \bibfnamefont [1]{#1}%
\providecommand \citenamefont [1]{#1}%
\providecommand \href@noop [0]{\@secondoftwo}%
\providecommand \href [0]{\begingroup \@sanitize@url \@href}%
\providecommand \@href[1]{\@@startlink{#1}\@@href}%
\providecommand \@@href[1]{\endgroup#1\@@endlink}%
\providecommand \@sanitize@url [0]{\catcode `\\12\catcode `\$12\catcode
  `\&12\catcode `\#12\catcode `\^12\catcode `\_12\catcode `\%12\relax}%
\providecommand \@@startlink[1]{}%
\providecommand \@@endlink[0]{}%
\providecommand \url  [0]{\begingroup\@sanitize@url \@url }%
\providecommand \@url [1]{\endgroup\@href {#1}{\urlprefix }}%
\providecommand \urlprefix  [0]{URL }%
\providecommand \Eprint [0]{\href }%
\providecommand \doibase [0]{https://doi.org/}%
\providecommand \selectlanguage [0]{\@gobble}%
\providecommand \bibinfo  [0]{\@secondoftwo}%
\providecommand \bibfield  [0]{\@secondoftwo}%
\providecommand \translation [1]{[#1]}%
\providecommand \BibitemOpen [0]{}%
\providecommand \bibitemStop [0]{}%
\providecommand \bibitemNoStop [0]{.\EOS\space}%
\providecommand \EOS [0]{\spacefactor3000\relax}%
\providecommand \BibitemShut  [1]{\csname bibitem#1\endcsname}%
\let\auto@bib@innerbib\@empty
\bibitem [{\citenamefont {Diosi}(2014)}]{Diosi2014}%
  \BibitemOpen
  \bibfield  {author} {\bibinfo {author} {\bibfnamefont {L.}~\bibnamefont
  {Di\'osi}},\ }\bibfield  {title} {\bibinfo {title} {{Hybrid quantum-classical
  master equations}},\ }\href
  {https://doi.org/10.1088/0031-8949/2014/T163/014004} {\bibfield  {journal}
  {\bibinfo  {journal} {Phys. Scr.}\ }\textbf {\bibinfo {volume} {T163}},\
  \bibinfo {pages} {14004} (\bibinfo {year} {2014})},\ \Eprint
  {https://arxiv.org/abs/1401.0476} {arXiv:1401.0476} \BibitemShut {NoStop}%
\bibitem [{\citenamefont {Buric}\ \emph {et~al.}(2013)\citenamefont {Buric},
  \citenamefont {Popovic}, \citenamefont {Radonjic},\ and\ \citenamefont
  {Prvanovic}}]{Buric2013}%
  \BibitemOpen
  \bibfield  {author} {\bibinfo {author} {\bibfnamefont {N.}~\bibnamefont
  {Buric}}, \bibinfo {author} {\bibfnamefont {D.~B.}\ \bibnamefont {Popovic}},
  \bibinfo {author} {\bibfnamefont {M.}~\bibnamefont {Radonjic}},\ and\
  \bibinfo {author} {\bibfnamefont {S.}~\bibnamefont {Prvanovic}},\ }\bibfield
  {title} {\bibinfo {title} {{Hybrid quantum-classical model of quantum
  measurements}},\ }\href {https://doi.org/10.1103/PhysRevA.87.054101}
  {\bibfield  {journal} {\bibinfo  {journal} {Phys. Rev. A}\ }\textbf {\bibinfo
  {volume} {87}},\ \bibinfo {pages} {54101} (\bibinfo {year}
  {2013})}\BibitemShut {NoStop}%
\bibitem [{\citenamefont {Martin-Dussaud}\ and\ \citenamefont
  {Rovelli}(2019)}]{Martin-Dussaud2019}%
  \BibitemOpen
  \bibfield  {author} {\bibinfo {author} {\bibfnamefont {P.}~\bibnamefont
  {Martin-Dussaud}}\ and\ \bibinfo {author} {\bibfnamefont {C.}~\bibnamefont
  {Rovelli}},\ }\bibfield  {title} {\bibinfo {title} {{Evaporating
  black-to-white hole}},\ }\href {https://doi.org/10.1088/1361-6382/ab5097}
  {\bibfield  {journal} {\bibinfo  {journal} {Classical Quantum Gravity}\
  }\textbf {\bibinfo {volume} {36}},\ \bibinfo {pages} {245002} (\bibinfo
  {year} {2019})},\ \Eprint {https://arxiv.org/abs/1905.07251v2}
  {arXiv:1905.07251v2} \BibitemShut {NoStop}%
\bibitem [{\citenamefont {Tilloy}(2019)}]{Tilloy2019}%
  \BibitemOpen
  \bibfield  {author} {\bibinfo {author} {\bibfnamefont {A.}~\bibnamefont
  {Tilloy}},\ }\bibfield  {title} {\bibinfo {title} {{Does gravity have to be
  quantized? Lessons from non-relativistic toy models}},\ }\href
  {https://doi.org/10.1088/1742-6596/1275/1/012006} {\bibfield  {journal}
  {\bibinfo  {journal} {J. Phys. Conf. Ser.}\ }\textbf {\bibinfo {volume}
  {1275}},\ \bibinfo {pages} {012006} (\bibinfo {year} {2019})},\ \Eprint
  {https://arxiv.org/abs/1903.01823} {arXiv:1903.01823} \BibitemShut {NoStop}%
\bibitem [{\citenamefont {C.~Tully}(1998)}]{Tully1998b}%
  \BibitemOpen
  \bibfield  {author} {\bibinfo {author} {\bibfnamefont {J.}~\bibnamefont
  {C.~Tully}},\ }\bibfield  {title} {\bibinfo {title} {Mixed
  quantum–classical dynamics},\ }\href {https://doi.org/10.1039/A801824C}
  {\bibfield  {journal} {\bibinfo  {journal} {Faraday Discuss.}\ }\textbf
  {\bibinfo {volume} {110}},\ \bibinfo {pages} {407} (\bibinfo {year}
  {1998})}\BibitemShut {NoStop}%
\bibitem [{\citenamefont {Yonehara}\ \emph {et~al.}(2012)\citenamefont
  {Yonehara}, \citenamefont {Hanasaki},\ and\ \citenamefont
  {Takatsuka}}]{Yonehara2012}%
  \BibitemOpen
  \bibfield  {author} {\bibinfo {author} {\bibfnamefont {T.}~\bibnamefont
  {Yonehara}}, \bibinfo {author} {\bibfnamefont {K.}~\bibnamefont {Hanasaki}},\
  and\ \bibinfo {author} {\bibfnamefont {K.}~\bibnamefont {Takatsuka}},\
  }\bibfield  {title} {\bibinfo {title} {{Fundamental Approaches to
  Nonadiabaticity: Toward a Chemical Theory beyond the Born–Oppenheimer
  Paradigm}},\ }\href {https://doi.org/10.1021/cr200096s} {\bibfield  {journal}
  {\bibinfo  {journal} {Chem. Rev.}\ }\textbf {\bibinfo {volume} {112}},\
  \bibinfo {pages} {499–542} (\bibinfo {year} {2012})}\BibitemShut {NoStop}%
\bibitem [{\citenamefont {Tavernelli}(2015)}]{Tavernelli2015}%
  \BibitemOpen
  \bibfield  {author} {\bibinfo {author} {\bibfnamefont {I.}~\bibnamefont
  {Tavernelli}},\ }\bibfield  {title} {\bibinfo {title} {Nonadiabatic molecular
  dynamics simulations: Synergies between theory and experiments},\ }\href
  {https://doi.org/10.1021/ar500357y} {\bibfield  {journal} {\bibinfo
  {journal} {Acc. Chem. Res.}\ }\textbf {\bibinfo {volume} {48}},\ \bibinfo
  {pages} {792} (\bibinfo {year} {2015})},\ \bibinfo {note} {pMID:
  25647401}\BibitemShut {NoStop}%
\bibitem [{\citenamefont {Crespo-Otero}\ and\ \citenamefont
  {Barbatti}(2018)}]{CrespoOtero2018}%
  \BibitemOpen
  \bibfield  {author} {\bibinfo {author} {\bibfnamefont {R.}~\bibnamefont
  {Crespo-Otero}}\ and\ \bibinfo {author} {\bibfnamefont {M.}~\bibnamefont
  {Barbatti}},\ }\bibfield  {title} {\bibinfo {title} {{Recent Advances and
  Perspectives on Nonadiabatic Mixed Quantum-Classical Dynamics}},\ }\href
  {https://doi.org/10.1021/acs.chemrev.7b00577} {\bibfield  {journal} {\bibinfo
   {journal} {Chem. Rev.}\ }\textbf {\bibinfo {volume} {118}},\ \bibinfo
  {pages} {7026–7068} (\bibinfo {year} {2018})}\BibitemShut {NoStop}%
\bibitem [{\citenamefont {Curchod}\ and\ \citenamefont
  {Martínez}(2018)}]{Curchod2018}%
  \BibitemOpen
  \bibfield  {author} {\bibinfo {author} {\bibfnamefont {B.~F.~E.}\
  \bibnamefont {Curchod}}\ and\ \bibinfo {author} {\bibfnamefont {T.~J.}\
  \bibnamefont {Mart{\'{\i}}nez}},\ }\bibfield  {title} {\bibinfo {title} {Ab initio
  nonadiabatic quantum molecular dynamics},\ }\href
  {https://doi.org/10.1021/acs.chemrev.7b00423} {\bibfield  {journal} {\bibinfo
   {journal} {Chem. Rev.}\ }\textbf {\bibinfo {volume} {118}},\ \bibinfo
  {pages} {3305} (\bibinfo {year} {2018})}\BibitemShut {NoStop}%
\bibitem [{\citenamefont {Prezhdo}\ and\ \citenamefont
  {Kisil}(1997)}]{Prezhdo1997}%
  \BibitemOpen
  \bibfield  {author} {\bibinfo {author} {\bibfnamefont {O.~V.}\ \bibnamefont
  {Prezhdo}}\ and\ \bibinfo {author} {\bibfnamefont {V.~V.}\ \bibnamefont
  {Kisil}},\ }\bibfield  {title} {\bibinfo {title} {{Mixing quantum and
  classical mechanics}},\ }\href
  {http://www.ams.org/mathscinet/search/publications.html?pg1=MR{\&}s1=MR1459700}
  {\bibfield  {journal} {\bibinfo  {journal} {Phys. Rev. A}\ }\textbf {\bibinfo
  {volume} {56}},\ \bibinfo {pages} {162} (\bibinfo {year}
  {1997})}\BibitemShut {NoStop}%
\bibitem [{\citenamefont {Kisil}(2005)}]{Kisil2005}%
  \BibitemOpen
  \bibfield  {author} {\bibinfo {author} {\bibfnamefont {V.~V.}\ \bibnamefont
  {Kisil}},\ }\bibfield  {title} {\bibinfo {title} {{A quantum-classical
  bracket from p -mechanics}},\ }\href
  {https://doi.org/10.1209/epl/i2005-10324-7} {\bibfield  {journal} {\bibinfo
  {journal} {Europhysics Letters (EPL)}\ }\textbf {\bibinfo {volume} {72}},\
  \bibinfo {pages} {873–879} (\bibinfo {year} {2005})}\BibitemShut {NoStop}%
\bibitem [{\citenamefont {Prezhdo}(2006)}]{Prezhdo2006}%
  \BibitemOpen
  \bibfield  {author} {\bibinfo {author} {\bibfnamefont {O.~V.}\ \bibnamefont
  {Prezhdo}},\ }\bibfield  {title} {\bibinfo {title} {{A quantum-classical
  bracket that satisfies the Jacobi identity.}},\ }\href
  {https://doi.org/10.1063/1.2200342} {\bibfield  {journal} {\bibinfo
  {journal} {J. Chem. Phys.}\ }\textbf {\bibinfo {volume} {124}},\ \bibinfo
  {pages} {201104} (\bibinfo {year} {2006})}\BibitemShut {NoStop}%
\bibitem [{\citenamefont {Salcedo}(2007)}]{Salcedo2007}%
  \BibitemOpen
  \bibfield  {author} {\bibinfo {author} {\bibfnamefont {L.~L.}\ \bibnamefont
  {Salcedo}},\ }\bibfield  {title} {\bibinfo {title} {Comment on ``{A}
  quantum-classical bracket that satisfies the {J}acobi identity'' [{J}.
  {C}hem. {P}hys. 124, 201104 (2006)]},\ }\href
  {https://doi.org/10.1063/1.2431650} {\bibfield  {journal} {\bibinfo
  {journal} {J. Chem. Phys.}\ }\textbf {\bibinfo {volume} {126}},\ \bibinfo
  {pages} {057101} (\bibinfo {year} {2007})},\ \Eprint
  {https://arxiv.org/abs/quant-ph/0701054v1} {quant-ph/0701054v1} \BibitemShut
  {NoStop}%
\bibitem [{\citenamefont {Agostini}\ \emph {et~al.}(2007)\citenamefont
  {Agostini}, \citenamefont {Caprara},\ and\ \citenamefont
  {Ciccotti}}]{Agostini2007}%
  \BibitemOpen
  \bibfield  {author} {\bibinfo {author} {\bibfnamefont {F.}~\bibnamefont
  {Agostini}}, \bibinfo {author} {\bibfnamefont {S.}~\bibnamefont {Caprara}},\
  and\ \bibinfo {author} {\bibfnamefont {G.}~\bibnamefont {Ciccotti}},\
  }\bibfield  {title} {\bibinfo {title} {{Do we have a consistent non-adiabatic
  quantum-classical mechanics?}},\ }\href
  {https://doi.org/10.1209/0295-5075/78/30001} {\bibfield  {journal} {\bibinfo
  {journal} {Europhysics Letters (EPL)}\ }\textbf {\bibinfo {volume} {78}},\
  \bibinfo {pages} {30001} (\bibinfo {year} {2007})}\BibitemShut {NoStop}%
\bibitem [{\citenamefont {Kisil}(2010)}]{Kisil2010}%
  \BibitemOpen
  \bibfield  {author} {\bibinfo {author} {\bibfnamefont {V.~V.}\ \bibnamefont
  {Kisil}},\ }\bibfield  {title} {\bibinfo {title} {{Comment on “Do we have a
  consistent non-adiabatic quantum-classical mechanics?” by Agostini F. et
  al.}},\ }\href {https://doi.org/10.1209/0295-5075/89/50005} {\bibfield
  {journal} {\bibinfo  {journal} {EPL (Europhysics Letters)}\ }\textbf
  {\bibinfo {volume} {89}},\ \bibinfo {pages} {50005} (\bibinfo {year}
  {2010})}\BibitemShut {NoStop}%
\bibitem [{\citenamefont {Agostini}\ \emph {et~al.}(2010)\citenamefont
  {Agostini}, \citenamefont {Caprara},\ and\ \citenamefont
  {Ciccotti}}]{Agostini2010}%
  \BibitemOpen
  \bibfield  {author} {\bibinfo {author} {\bibfnamefont {F.}~\bibnamefont
  {Agostini}}, \bibinfo {author} {\bibfnamefont {S.}~\bibnamefont {Caprara}},\
  and\ \bibinfo {author} {\bibfnamefont {G.}~\bibnamefont {Ciccotti}},\
  }\bibfield  {title} {\bibinfo {title} {{Reply to the Comment by VV Kisil}},\
  }\href
  {http://iopscience.iop.org/0295-5075/89/5/50006/pdf/0295-5075{\_}89{\_}5{\_}50006.pdf}
  {\bibfield  {journal} {\bibinfo  {journal} {EPL (Europhysics Letters)}\
  }\textbf {\bibinfo {volume} {89}},\ \bibinfo {pages} {50006} (\bibinfo {year}
  {2010})}\BibitemShut {NoStop}%
\bibitem [{\citenamefont {Hall}(2008)}]{Hall2008}%
  \BibitemOpen
  \bibfield  {author} {\bibinfo {author} {\bibfnamefont {M.~J.~W.}\
  \bibnamefont {Hall}},\ }\bibfield  {title} {\bibinfo {title} {{Consistent
  classical and quantum mixed dynamics}},\ }\href
  {https://doi.org/10.1103/PhysRevA.78.042104} {\bibfield  {journal} {\bibinfo
  {journal} {Phys. Rev. A}\ }\textbf {\bibinfo {volume} {78}},\ \bibinfo
  {pages} {42104} (\bibinfo {year} {2008})},\ \Eprint
  {https://arxiv.org/abs/0804.2505} {arXiv:0804.2505} \BibitemShut {NoStop}%
\bibitem [{\citenamefont {Buric}\ \emph {et~al.}(2013)\citenamefont {Buric},
  \citenamefont {Popovic}, \citenamefont {Radonjic},\ and\ \citenamefont
  {Prvanovic}}]{Buric2013b}%
  \BibitemOpen
  \bibfield  {author} {\bibinfo {author} {\bibfnamefont {N.}~\bibnamefont
  {Burić}}, \bibinfo {author} {\bibfnamefont {D.~B.}\ \bibnamefont
  {Popović}}, \bibinfo {author} {\bibfnamefont {M.}~\bibnamefont
  {Radonjić}},\ and\ \bibinfo {author} {\bibfnamefont {S.}~\bibnamefont
  {Prvanović}},\ }\bibfield  {title} {\bibinfo {title} {{Hamiltonian
  Formulation of Statistical Ensembles and Mixed States of Quantum and Hybrid
  Systems}},\ }\href {https://doi.org/10.1007/s10701-013-9755-z} {\bibfield
  {journal} {\bibinfo  {journal} {Found. Phys.}\ }\textbf {\bibinfo {volume}
  {43}},\ \bibinfo {pages} {1459–1477} (\bibinfo {year} {2013})}\BibitemShut
  {NoStop}%
\bibitem [{\citenamefont {Peres}\ and\ \citenamefont
  {Terno}(2001)}]{Peres2001}%
  \BibitemOpen
  \bibfield  {author} {\bibinfo {author} {\bibfnamefont {A.}~\bibnamefont
  {Peres}}\ and\ \bibinfo {author} {\bibfnamefont {D.~R.}\ \bibnamefont
  {Terno}},\ }\bibfield  {title} {\bibinfo {title} {Hybrid classical-quantum
  dynamics},\ }\href {https://doi.org/10.1103/physreva.63.022101} {\bibfield
  {journal} {\bibinfo  {journal} {Physical Review A}\ }\textbf {\bibinfo
  {volume} {63}},\ \bibinfo {pages} {022101} (\bibinfo {year}
  {2001})}\BibitemShut {NoStop}%
\bibitem [{\citenamefont {Terno}(2006)}]{Terno2006}%
  \BibitemOpen
  \bibfield  {author} {\bibinfo {author} {\bibfnamefont {D.~R.}\ \bibnamefont
  {Terno}},\ }\bibfield  {title} {\bibinfo {title} {Inconsistency of
  quantum--classical dynamics, and what it implies},\ }\href
  {https://doi.org/10.1007/s10701-005-9007-y} {\bibfield  {journal} {\bibinfo
  {journal} {Found. Phys.}\ }\textbf {\bibinfo {volume} {36}},\ \bibinfo
  {pages} {102} (\bibinfo {year} {2006})},\ \Eprint
  {https://arxiv.org/abs/quant-ph/0402092v1} {quant-ph/0402092v1} \BibitemShut
  {NoStop}%
\bibitem [{\citenamefont {Salcedo}(1996)}]{Salcedo1996}%
  \BibitemOpen
  \bibfield  {author} {\bibinfo {author} {\bibfnamefont {L.~L.}\ \bibnamefont
  {Salcedo}},\ }\bibfield  {title} {\bibinfo {title} {Absence of classical and
  quantum mixing},\ }\href {https://doi.org/10.1103/PhysRevA.54.3657}
  {\bibfield  {journal} {\bibinfo  {journal} {Phys. Rev A}\ }\textbf {\bibinfo
  {volume} {54}},\ \bibinfo {pages} {3657} (\bibinfo {year} {1996})},\ \Eprint
  {https://arxiv.org/abs/hep-th/9509089v1} {hep-th/9509089v1} \BibitemShut
  {NoStop}%
\bibitem [{\citenamefont {Gil}\ and\ \citenamefont {Salcedo}(2017)}]{Gil2017}%
  \BibitemOpen
  \bibfield  {author} {\bibinfo {author} {\bibfnamefont {V.}~\bibnamefont
  {Gil}}\ and\ \bibinfo {author} {\bibfnamefont {L.~L.}\ \bibnamefont
  {Salcedo}},\ }\bibfield  {title} {\bibinfo {title} {{Canonical bracket in
  quantum-classical hybrid systems}},\ }\href
  {https://doi.org/10.1103/PhysRevA.95.012137} {\bibfield  {journal} {\bibinfo
  {journal} {Phys. Rev. A}\ }\textbf {\bibinfo {volume} {95}},\ \bibinfo
  {pages} {012137} (\bibinfo {year} {2017})},\ \Eprint
  {https://arxiv.org/abs/1612.05799} {arXiv:1612.05799} \BibitemShut {NoStop}%
\bibitem [{\citenamefont {Caro}\ and\ \citenamefont
  {Salcedo}(1999)}]{Caro1999}%
  \BibitemOpen
  \bibfield  {author} {\bibinfo {author} {\bibfnamefont {J.}~\bibnamefont
  {Caro}}\ and\ \bibinfo {author} {\bibfnamefont {L.~L.}\ \bibnamefont
  {Salcedo}},\ }\bibfield  {title} {\bibinfo {title} {Impediments to mixing
  classical and quantum dynamics},\ }\href
  {https://doi.org/10.1103/PhysRevA.60.842} {\bibfield  {journal} {\bibinfo
  {journal} {Phys. Rev. A}\ }\textbf {\bibinfo {volume} {60}},\ \bibinfo
  {pages} {842} (\bibinfo {year} {1999})}\BibitemShut {NoStop}%
\bibitem [{\citenamefont {Elze}(2012)}]{Elze2012}%
  \BibitemOpen
  \bibfield  {author} {\bibinfo {author} {\bibfnamefont {H.}~\bibnamefont
  {Elze}},\ }\bibfield  {title} {\bibinfo {title} {{Linear dynamics of
  quantum-classical hybrids}},\ }\href
  {https://doi.org/10.1103/PhysRevA.85.052109} {\bibfield  {journal} {\bibinfo
  {journal} {Phys. Rev. A}\ }\textbf {\bibinfo {volume} {85}},\ \bibinfo
  {pages} {52109} (\bibinfo {year} {2012})},\ \Eprint
  {https://arxiv.org/abs/1111.2276} {arXiv:1111.2276} \BibitemShut {NoStop}%
\bibitem [{\citenamefont {Aleksandrov}(1981)}]{Aleksandrov1981}%
  \BibitemOpen
  \bibfield  {author} {\bibinfo {author} {\bibfnamefont {I.~V.}\ \bibnamefont
  {Aleksandrov}},\ }\bibfield  {title} {\bibinfo {title} {The statistical
  dynamics of a system consisting of a classical and a quantum system},\
  }\href@noop {} {\bibfield  {journal} {\bibinfo  {journal} {Z. Naturforsch}\
  }\textbf {\bibinfo {volume} {36a}},\ \bibinfo {pages} {902} (\bibinfo {year}
  {1981})}\BibitemShut {NoStop}%
\bibitem [{\citenamefont {Kapral}\ and\ \citenamefont
  {Ciccotti}(1999)}]{Kapral1999}%
  \BibitemOpen
  \bibfield  {author} {\bibinfo {author} {\bibfnamefont {R.}~\bibnamefont
  {Kapral}}\ and\ \bibinfo {author} {\bibfnamefont {G.}~\bibnamefont
  {Ciccotti}},\ }\bibfield  {title} {\bibinfo {title} {{Mixed quantum-classical
  dynamics}},\ }\href {http://link.aip.org/link/?JCPSA6/110/8919/1} {\bibfield
  {journal} {\bibinfo  {journal} {J. Chem. Phys.}\ }\textbf {\bibinfo {volume}
  {110}},\ \bibinfo {pages} {8919–8929} (\bibinfo {year} {1999})}\BibitemShut
  {NoStop}%
\bibitem [{\citenamefont {Alonso}\ \emph {et~al.}(2011)\citenamefont {Alonso},
  \citenamefont {Castro}, \citenamefont {Clemente-Gallardo}, \citenamefont
  {Cuch{\'{\i}}}, \citenamefont {Echenique},\ and\ \citenamefont
  {Falceto}}]{Alonso2011}%
  \BibitemOpen
  \bibfield  {author} {\bibinfo {author} {\bibfnamefont {J.~L.}\ \bibnamefont
  {Alonso}}, \bibinfo {author} {\bibfnamefont {A.}~\bibnamefont {Castro}},
  \bibinfo {author} {\bibfnamefont {J.}~\bibnamefont {Clemente-Gallardo}},
  \bibinfo {author} {\bibfnamefont {J.~C.}\ \bibnamefont {Cuch{\'{\i}}}}, \bibinfo
  {author} {\bibfnamefont {P.}~\bibnamefont {Echenique}},\ and\ \bibinfo
  {author} {\bibfnamefont {F.}~\bibnamefont {Falceto}},\ }\bibfield  {title}
  {\bibinfo {title} {{Statistics and Nosé formalism for Ehrenfest dynamics}},\
  }\href {https://doi.org/10.1088/1751-8113/44/39/395004} {\bibfield  {journal}
  {\bibinfo  {journal} {J. Phys. A: Math. Theor.}\ }\textbf {\bibinfo {volume}
  {44}},\ \bibinfo {pages} {395004} (\bibinfo {year} {2011})}\BibitemShut
  {NoStop}%
\bibitem [{\citenamefont {Alavi}\ \emph {et~al.}(1994)\citenamefont {Alavi},
  \citenamefont {Kohanoff}, \citenamefont {Parrinello},\ and\ \citenamefont
  {Frenkel}}]{Alavi1994}%
  \BibitemOpen
  \bibfield  {author} {\bibinfo {author} {\bibfnamefont {A.}~\bibnamefont
  {Alavi}}, \bibinfo {author} {\bibfnamefont {J.}~\bibnamefont {Kohanoff}},
  \bibinfo {author} {\bibfnamefont {M.}~\bibnamefont {Parrinello}},\ and\
  \bibinfo {author} {\bibfnamefont {D.}~\bibnamefont {Frenkel}},\ }\bibfield
  {title} {\bibinfo {title} {{Ab Initio molecular dynamics with excited
  electrons}},\ }\href {https://doi.org/10.1103/PhysRevLett.73.2599} {\bibfield
   {journal} {\bibinfo  {journal} {Phys. Rev. Lett.}\ }\textbf {\bibinfo
  {volume} {73}},\ \bibinfo {pages} {2599} (\bibinfo {year}
  {1994})}\BibitemShut {NoStop}%
\bibitem [{\citenamefont {Grumbach}\ \emph {et~al.}(1994)\citenamefont
  {Grumbach}, \citenamefont {Hohl}, \citenamefont {Martin},\ and\ \citenamefont
  {Car}}]{Grumbach1994}%
  \BibitemOpen
  \bibfield  {author} {\bibinfo {author} {\bibfnamefont {M.~P.}\ \bibnamefont
  {Grumbach}}, \bibinfo {author} {\bibfnamefont {D.}~\bibnamefont {Hohl}},
  \bibinfo {author} {\bibfnamefont {R.~M.}\ \bibnamefont {Martin}},\ and\
  \bibinfo {author} {\bibfnamefont {R.}~\bibnamefont {Car}},\ }\bibfield
  {title} {\bibinfo {title} {{Ab initio molecular dynamics with a
  finite-temperature density functional}},\ }\href
  {https://doi.org/10.1088/0953-8984/6/10/017} {\bibfield  {journal} {\bibinfo
  {journal} {J. Phys.: Condens. Matter}\ }\textbf {\bibinfo {volume} {6}},\
  \bibinfo {pages} {1999–2014} (\bibinfo {year} {1994})}\BibitemShut
  {NoStop}%
\bibitem [{\citenamefont {Silvestrelli}\ \emph {et~al.}(1996)\citenamefont
  {Silvestrelli}, \citenamefont {Alavi}, \citenamefont {Parrinello},\ and\
  \citenamefont {Frenkel}}]{Silvestrelli1996}%
  \BibitemOpen
  \bibfield  {author} {\bibinfo {author} {\bibfnamefont {P.~L.}\ \bibnamefont
  {Silvestrelli}}, \bibinfo {author} {\bibfnamefont {A.}~\bibnamefont {Alavi}},
  \bibinfo {author} {\bibfnamefont {M.}~\bibnamefont {Parrinello}},\ and\
  \bibinfo {author} {\bibfnamefont {D.}~\bibnamefont {Frenkel}},\ }\bibfield
  {title} {\bibinfo {title} {{Ab initio molecular dynamics simulation of laser
  melting of silicon}},\ }\href {https://doi.org/10.1103/PhysRevLett.77.3149}
  {\bibfield  {journal} {\bibinfo  {journal} {Phys. Rev. Lett.}\ }\textbf
  {\bibinfo {volume} {77}},\ \bibinfo {pages} {3149} (\bibinfo {year}
  {1996})}\BibitemShut {NoStop}%
\bibitem [{\citenamefont {Ji}\ and\ \citenamefont {Zhang}(2013)}]{Ji2013}%
  \BibitemOpen
  \bibfield  {author} {\bibinfo {author} {\bibfnamefont {P.}~\bibnamefont
  {Ji}}\ and\ \bibinfo {author} {\bibfnamefont {Y.}~\bibnamefont {Zhang}},\
  }\bibfield  {title} {\bibinfo {title} {{Femtosecond laser processing of
  germanium: An ab initio molecular dynamics study}},\ }\href
  {https://doi.org/10.1088/0022-3727/46/49/495108} {\bibfield  {journal}
  {\bibinfo  {journal} {J. Phys. D: Appl. Phys.}\ }\textbf {\bibinfo {volume}
  {46}},\ \bibinfo {pages} {495108} (\bibinfo {year} {2013})}\BibitemShut
  {NoStop}%
\bibitem [{\citenamefont {R\"uter}\ and\ \citenamefont
  {Redmer}(2014)}]{Ruter2014}%
  \BibitemOpen
  \bibfield  {author} {\bibinfo {author} {\bibfnamefont {H.~R.}\ \bibnamefont
  {R\"uter}}\ and\ \bibinfo {author} {\bibfnamefont {R.}~\bibnamefont
  {Redmer}},\ }\bibfield  {title} {\bibinfo {title} {{Ab Initio Simulations for
  the Ion-Ion Structure Factor of Warm Dense Aluminum}},\ }\href
  {https://doi.org/10.1103/PhysRevLett.112.145007} {\bibfield  {journal}
  {\bibinfo  {journal} {Phys. Rev. Lett.}\ }\textbf {\bibinfo {volume} {112}},\
  \bibinfo {pages} {145007} (\bibinfo {year} {2014})}\BibitemShut {NoStop}%
\bibitem [{\citenamefont {Karasiev}\ \emph {et~al.}(2014)\citenamefont
  {Karasiev}, \citenamefont {Sjostrom},\ and\ \citenamefont
  {Trickey}}]{Karasiev2014}%
  \BibitemOpen
  \bibfield  {author} {\bibinfo {author} {\bibfnamefont {V.~V.}\ \bibnamefont
  {Karasiev}}, \bibinfo {author} {\bibfnamefont {T.}~\bibnamefont {Sjostrom}},\
  and\ \bibinfo {author} {\bibfnamefont {S.~B.}\ \bibnamefont {Trickey}},\
  }\bibfield  {title} {\bibinfo {title} {{Finite-temperature orbital-free DFT
  molecular dynamics: Coupling Profess and Quantum Espresso}},\ }\href
  {https://doi.org/10.1016/j.cpc.2014.08.023} {\bibfield  {journal} {\bibinfo
  {journal} {Comput. Phys. Commun.}\ }\textbf {\bibinfo {volume} {185}},\
  \bibinfo {pages} {3240–3249} (\bibinfo {year} {2014})},\ \Eprint
  {https://arxiv.org/abs/1406.0835} {arXiv:1406.0835} \BibitemShut {NoStop}%
\bibitem [{\citenamefont {{Von Neumann}}(1955)}]{vonNeumann1955}%
  \BibitemOpen
  \bibfield  {author} {\bibinfo {author} {\bibfnamefont {J.}~\bibnamefont {{Von
  Neumann}}},\ }\href@noop {} {\emph {\bibinfo {title} {{Mathematical
  foundations of quantum mechanics}}}}\ (\bibinfo  {publisher} {Princeton
  University Press, Princeton},\ \bibinfo {year} {1955})\BibitemShut {NoStop}%
\bibitem [{\citenamefont {Bornemann}\ \emph {et~al.}(1996)\citenamefont
  {Bornemann}, \citenamefont {Nettesheim},\ and\ \citenamefont
  {Schütte}}]{Bornemann1996}%
  \BibitemOpen
  \bibfield  {author} {\bibinfo {author} {\bibfnamefont {F.~A.~A.}\
  \bibnamefont {Bornemann}}, \bibinfo {author} {\bibfnamefont {P.}~\bibnamefont
  {Nettesheim}},\ and\ \bibinfo {author} {\bibfnamefont {C.}~\bibnamefont
  {Schütte}},\ }\bibfield  {title} {\bibinfo {title} {{Quantum-classical
  molecular dynamics as an approximation to full quantum dynamics}},\ }\href
  {http://www.zib.eu/Publications/Reports/SC-95-26.pdf} {\bibfield  {journal}
  {\bibinfo  {journal} {J. Chem. Phys,}\ }\textbf {\bibinfo {volume} {105}},\
  \bibinfo {pages} {1074–1083} (\bibinfo {year} {1996})}\BibitemShut
  {NoStop}%
\bibitem [{\citenamefont {Kibble}(1979)}]{Kibble1979}%
  \BibitemOpen
  \bibfield  {author} {\bibinfo {author} {\bibfnamefont {T.}~\bibnamefont
  {Kibble}},\ }\bibfield  {title} {\bibinfo {title} {{Geometrization of quantum
  mechanics}},\ }\href {http://www.springerlink.com/index/N3867L012385K347.pdf}
  {\bibfield  {journal} {\bibinfo  {journal} {Commun. Math. Phys.}\ }\textbf
  {\bibinfo {volume} {65}},\ \bibinfo {pages} {189–201} (\bibinfo {year}
  {1979})}\BibitemShut {NoStop}%
\bibitem [{\citenamefont {Heslot}(1985)}]{Heslot1985}%
  \BibitemOpen
  \bibfield  {author} {\bibinfo {author} {\bibfnamefont {A.}~\bibnamefont
  {Heslot}},\ }\bibfield  {title} {\bibinfo {title} {{Quantum mechanics as a
  classical theory}},\ }\href
  {http://link.aps.org/doi/10.1103/PhysRevD.31.1341} {\bibfield  {journal}
  {\bibinfo  {journal} {Phys. Rev. D}\ }\textbf {\bibinfo {volume} {31}},\
  \bibinfo {pages} {1341} (\bibinfo {year} {1985})}\BibitemShut {NoStop}%
\bibitem [{\citenamefont {Brody}\ and\ \citenamefont
  {Hughston}(1998)}]{Brody1998}%
  \BibitemOpen
  \bibfield  {author} {\bibinfo {author} {\bibfnamefont {D.~C.}\ \bibnamefont
  {Brody}}\ and\ \bibinfo {author} {\bibfnamefont {L.~P.}\ \bibnamefont
  {Hughston}},\ }\bibfield  {title} {\bibinfo {title} {{The quantum canonical
  ensemble}},\ }\href {https://doi.org/10.1063/1.532661} {\bibfield  {journal}
  {\bibinfo  {journal} {J. Math. Phys.}\ }\textbf {\bibinfo {volume} {39}},\
  \bibinfo {pages} {6502–6508} (\bibinfo {year} {1998})}\BibitemShut
  {NoStop}%
\bibitem [{\citenamefont {Jona-Lasinio}\ and\ \citenamefont
  {Presilla}(2006)}]{JonaLasinio2006}%
  \BibitemOpen
  \bibfield  {author} {\bibinfo {author} {\bibfnamefont {G.}~\bibnamefont
  {Jona-Lasinio}}\ and\ \bibinfo {author} {\bibfnamefont {C.}~\bibnamefont
  {Presilla}},\ }\bibfield  {title} {\bibinfo {title} {{On the statistics of
  quantum expectations for systems in thermal equilibrium}},\ }in\ \href@noop
  {} {\emph {\bibinfo {booktitle} {{AIP Conf. Proc. 844}}}},\ Vol.\ \bibinfo
  {volume} {844}\ (\bibinfo  {publisher} {AIP},\ \bibinfo {year} {2006})\ p.\
  \bibinfo {pages} {200–205}\BibitemShut {NoStop}%
\bibitem [{\citenamefont {Campisi}(2013)}]{Campisi2013a}%
  \BibitemOpen
  \bibfield  {author} {\bibinfo {author} {\bibfnamefont {M.}~\bibnamefont
  {Campisi}},\ }\bibfield  {title} {\bibinfo {title} {{Quantum fluctuation
  relations for ensembles of wave functions}},\ }\href
  {https://doi.org/10.1088/1367-2630/15/11/115008} {\bibfield  {journal}
  {\bibinfo  {journal} {New J. Phys.}\ }\textbf {\bibinfo {volume} {15}},\
  \bibinfo {pages} {115008} (\bibinfo {year} {2013})}\BibitemShut {NoStop}%
\bibitem [{\citenamefont {Alonso}\ \emph {et~al.}(2015)\citenamefont {Alonso},
  \citenamefont {Castro}, \citenamefont {Clemente-Gallardo}, \citenamefont
  {Cuch{\'{\i}}}, \citenamefont {Echenique}, \citenamefont {Esteve},\ and\
  \citenamefont {Falceto}}]{Alonso2015}%
  \BibitemOpen
  \bibfield  {author} {\bibinfo {author} {\bibfnamefont {J.~L.}\ \bibnamefont
  {Alonso}}, \bibinfo {author} {\bibfnamefont {A.}~\bibnamefont {Castro}},
  \bibinfo {author} {\bibfnamefont {J.}~\bibnamefont {Clemente-Gallardo}},
  \bibinfo {author} {\bibfnamefont {J.~C.}\ \bibnamefont {Cuch{\'{\i}}}},
  \bibinfo {author} {\bibfnamefont {P.}~\bibnamefont {Echenique}}, \bibinfo
  {author} {\bibfnamefont {J.~G.}\ \bibnamefont {Esteve}},\ and\ \bibinfo
  {author} {\bibfnamefont {F.}~\bibnamefont {Falceto}},\ }\bibfield  {title}
  {\bibinfo {title} {{Nonextensive thermodynamic functions in the
  Schrödinger-Gibbs ensemble}},\ }\href
  {https://doi.org/10.1103/PhysRevE.91.022137} {\bibfield  {journal} {\bibinfo
  {journal} {Phys. Rev. E}\ }\textbf {\bibinfo {volume} {91}},\ \bibinfo
  {pages} {022137} (\bibinfo {year} {2015})}\BibitemShut {NoStop}%
\bibitem [{\citenamefont {Gleason}(1957)}]{Gleason1957}%
  \BibitemOpen
  \bibfield  {author} {\bibinfo {author} {\bibfnamefont {A.~M.}\ \bibnamefont
  {Gleason}},\ }\bibfield  {title} {\bibinfo {title} {{Measures on the closed
  subspaces of a Hilbert space}},\ }\href
  {http://books.google.com/books?hl=en{\&}lr={\&}id=BpvLhvMfmagC{\&}oi=fnd{\&}pg=PA123{\&}dq=Measures+on+the+closed+subspaces+of+a+Hilbert+space{\&}ots=ZEoxey99tF{\&}sig=07zB1IQZvlexj3QxEFfkNd5pn28}
  {\bibfield  {journal} {\bibinfo  {journal} {J. of Mathematics and Mechanics}\
  }\textbf {\bibinfo {volume} {6}},\ \bibinfo {pages} {885–893} (\bibinfo
  {year} {1957})}\BibitemShut {NoStop}%
\bibitem [{\citenamefont {Busch}(2003)}]{Busch2003}%
  \BibitemOpen
  \bibfield  {author} {\bibinfo {author} {\bibfnamefont {P.}~\bibnamefont
  {Busch}},\ }\bibfield  {title} {\bibinfo {title} {{Quantum States and
  Generalized Observables: A Simple Proof of Gleason's Theorem}},\ }\href
  {https://doi.org/10.1103/PhysRevLett.91.120403} {\bibfield  {journal}
  {\bibinfo  {journal} {Phys. Rev. Lett.}\ }\textbf {\bibinfo {volume} {91}},\
  \bibinfo {pages} {120403} (\bibinfo {year} {2003})}\BibitemShut {NoStop}%
\bibitem [{\citenamefont {Caves}\ \emph {et~al.}(2004)\citenamefont {Caves},
  \citenamefont {Fuchs}, \citenamefont {Manne},\ and\ \citenamefont
  {Renes}}]{Caves2004}%
  \BibitemOpen
  \bibfield  {author} {\bibinfo {author} {\bibfnamefont {C.~M.}\ \bibnamefont
  {Caves}}, \bibinfo {author} {\bibfnamefont {C.~A.}\ \bibnamefont {Fuchs}},
  \bibinfo {author} {\bibfnamefont {K.~K.}\ \bibnamefont {Manne}},\ and\
  \bibinfo {author} {\bibfnamefont {J.~M.}\ \bibnamefont {Renes}},\ }\bibfield
  {title} {\bibinfo {title} {{Gleason-type derivations of the quantum
  probability rule for generalized measurements}},\ }\href
  {https://doi.org/10.1023/B:FOOP.0000019581.00318.a5} {\bibfield  {journal}
  {\bibinfo  {journal} {Found. Phys.}\ }\textbf {\bibinfo {volume} {34}},\
  \bibinfo {pages} {193–209} (\bibinfo {year} {2004})},\ \Eprint
  {https://arxiv.org/abs/0306179} {arXiv:0306179 [quant-ph]} \BibitemShut
  {NoStop}%
\bibitem [{Note1()}]{Note1}%
  \BibitemOpen
  \bibinfo {note} {For example, it was given in Ref.~\cite {Mauri1993}, where
  it was claimed to be the partial classical limit of the fully quantum
  canonical ensemble. It was also presented as the zero-th order term in a
  classical-limit expansion of the partial Wigner transformation of the quantum
  canonical ensemble in Ref.~\cite {Kapral1999}. Finally, in footnote 30 of
  Ref.~\cite {Alonso2012c}, some of the current authors already hinted, without
  proof, the result demonstrated here.}\BibitemShut {Stop}%
\bibitem [{\citenamefont {Klein}(1931)}]{Klein1931}%
  \BibitemOpen
  \bibfield  {author} {\bibinfo {author} {\bibfnamefont {O.}~\bibnamefont
  {Klein}},\ }\bibfield  {title} {\bibinfo {title} {Zur quantenmechanischen
  {B}egr{\"u}ndung des zweiten {H}auptsatzes der {W}{\"a}rmelehre},\ }\href
  {https://doi.org/10.1007/BF01341997} {\bibfield  {journal} {\bibinfo
  {journal} {Zeitschrift f{\"u}r Physik}\ }\textbf {\bibinfo {volume} {72}},\
  \bibinfo {pages} {767} (\bibinfo {year} {1931})}\BibitemShut {NoStop}%
\bibitem [{\citenamefont {Mauri}\ \emph {et~al.}(1993)\citenamefont {Mauri},
  \citenamefont {Car},\ and\ \citenamefont {Tosatti}}]{Mauri1993}%
  \BibitemOpen
  \bibfield  {author} {\bibinfo {author} {\bibfnamefont {F.}~\bibnamefont
  {Mauri}}, \bibinfo {author} {\bibfnamefont {R.}~\bibnamefont {Car}},\ and\
  \bibinfo {author} {\bibfnamefont {E.}~\bibnamefont {Tosatti}},\ }\bibfield
  {title} {\bibinfo {title} {{Canonical Statistical Averages of Coupled}},\
  }\href {https://doi.org/10.1209/0295-5075/24/6/001} {\bibfield  {journal}
  {\bibinfo  {journal} {Europhysics Letters (EPL)}\ }\textbf {\bibinfo {volume}
  {24}},\ \bibinfo {pages} {431–436} (\bibinfo {year} {1993})}\BibitemShut
  {NoStop}%
\bibitem [{\citenamefont {Alonso}\ \emph {et~al.}(2012)\citenamefont {Alonso},
  \citenamefont {Castro}, \citenamefont {Clemente-Gallardo}, \citenamefont
  {Echenique}, \citenamefont {Mazo}, \citenamefont {Polo}, \citenamefont
  {Rubio},\ and\ \citenamefont {Zueco}}]{Alonso2012c}%
  \BibitemOpen
  \bibfield  {author} {\bibinfo {author} {\bibfnamefont {J.~L.}\ \bibnamefont
  {Alonso}}, \bibinfo {author} {\bibfnamefont {A.}~\bibnamefont {Castro}},
  \bibinfo {author} {\bibfnamefont {J.}~\bibnamefont {Clemente-Gallardo}},
  \bibinfo {author} {\bibfnamefont {P.}~\bibnamefont {Echenique}}, \bibinfo
  {author} {\bibfnamefont {J.~J.}\ \bibnamefont {Mazo}}, \bibinfo {author}
  {\bibfnamefont {V.}~\bibnamefont {Polo}}, \bibinfo {author} {\bibfnamefont
  {A.}~\bibnamefont {Rubio}},\ and\ \bibinfo {author} {\bibfnamefont
  {D.}~\bibnamefont {Zueco}},\ }\bibfield  {title} {\bibinfo {title}
  {{Non-adiabatic effects within a single thermally averaged potential energy
  surface: Thermal expansion and reaction rates of small molecules}},\ }\href
  {https://doi.org/10.1063/1.4747699} {\bibfield  {journal} {\bibinfo
  {journal} {J. Chem. Phys.}\ }\textbf {\bibinfo {volume} {137}},\ \bibinfo
  {pages} {22A533} (\bibinfo {year} {2012})}\BibitemShut {NoStop}%
\end{thebibliography}
%

\end{document}